# Open Source-Based Over-the-Air 5G New Radio Sidelink Testbed

Melissa Elkadi   Deokseong Kim   Ejaz Ahmed   Anh Le   Moein Sadeghi   Paul Russell   Bo Ryu

EpiSys Science, Inc., Poway, CA, USA

{melissa, david.kim, ejaz, moein, anhle, paul.russell, bo.ryu}@episci.com

*Abstract*— **The focus of this paper is to demonstrate an over-the-air (OTA) 5G new radio (NR) sidelink communication prototype. 5G NR sidelink communications allow NR UEs to transfer data independently without the assistance of a base station (gNB), which enables V2X communications, including platooning, autonomous driving, sensor extension, industrial IoT, public safety communication and much more. Our design leverages the open-source OpenAirInterface5G (OAI) software, which operates on software-defined radios (SDRs) and can be easily extended for mesh networking. The software includes all signal processing components specified by the 3GPP 5G sidelink standards, including Low-Density Parity Check (LDPC) encoding/decoding, polar encoding/decoding, data and control multiplexing, modulation/demodulation, and orthogonal frequency-division multiplexing (OFDM) modulation/demodulation. It can be configured to operate with different bands, bandwidths, and antenna settings.**

**The first milestone in this work was to demonstrate the completed Physical Sidelink Broadcast Channel (PSBCH) development, which conducts synchronization between a Synchronization Reference (SyncRef) UE and a nearby UE. The SyncRef UE broadcasts a sidelink synchronization signal block (S-SSB) periodically, which the nearby UE detects and uses to synchronize its timing and frequency components with the SyncRef UE. Once a connection is established, the next developmental milestone is to transmit real data (text messages) via the Physical Sidelink Shared Channel (PSSCH). Our PHY sidelink framework is tested using both an RF simulator and an OTA testbed with multiple nearby UEs. Beyond the development of synchronization and data transmission/reception in 5G sidelink, we conclude with various performance tests and validation experiments. The results of these metrics show that our simulator is comparable to the OTA testbed and can be used for upper layer development in the future.**

I. INTRODUCTION

As 5G NR becomes more and more widespread, there is a critical need to research and study the possible use cases and potential performance of emerging 5G standards, systems, solutions, and their adaptations under a wide range of scenarios [1]. One underdeveloped but highly sought after feature of emerging 5G technologies is 5G Sidelink (SL) [2], [3]. 5G SL enables device-to-device (D2D) communications supporting data transfer with and without the assistance of a base station (gNB). SL has a large spectrum of use cases and applications. The use of SL technology can be applied to any environment requiring efficient and reliable communication, V2X communications, machine learning (ML), reinforcement learning (RL), mesh and relay applications, and many more. SL is designed to provide a comprehensive set of advantages to the overall 5G system, as it can operate in various spectrum configurations, including dedicated in-band licensed and unlicensed ranges, allowing it to be deployed in various environments. Furthermore, it can support a broad range of devices, enabling a diverse ecosystem of use cases. There are several benefits and applications to 5G SL technology, including enhancing military communications through:

- vehicle platooning,
- advanced autonomous driving,
- increased sensor extension,
- remote driving capabilities,
- coverage extension,
- location tracking enhancements in various adverse environments, and
- data offloading speeds.

Other domain applications include the improvement of transportation and public safety communication, expanding the industrial IoT system capacity, and improving wearable connectivity applications [4].

Given that 5G SL is currently not available in the open-source community, this paper will document the benefits of 5G SL, details of the Physical (PHY) layer SL channels, the approaches and assumptions made in the development cycle, and the results of both the SL synchronization procedure and the transmission and successful reception of data packets over-the-air (OTA) between multiple software-defined radios (SDRs). All of the work described in this paper was conducted only for SL mode 2; meaning there is no gNB presence and the communication is D2D. Our contributions in this work can be summarized as follows:

- **Analysis**: A detailed analysis was conducted over all physical SL channels and signals, including physical SL broadcast channel (PSBCH), shared channel (PSSCH), control channel (PSCCH), feedback channel (PSFCH), and SL primary and secondary synchronization signals (SPSS and SSSS).
- **Development**: Using OpenAirInterface5G Software (OAI) [5], a comprehensive physical layer communication framework for 5G SL communication was developed, with the emphasis on mode 2 (D2D). Our technical paradigm follows 3GPP Release 16 standard TS 38.211 to 38.214 for 5G SL.
- **Validation**: The development framework was validated and stressed with added white gaussian noise (AWGN) via an RF simulator (RFSIM). RFSIM allows for both functional and performance verification metrics to be gathered and evaluated. Reference signal received power (RSRP) and block error rate (BLER) measurements we recorded for both synchronization and data transfer procedures. These measurements were taken at various SNR levels to support claims of true 5G SL functionality.
- **Verification**: Finally, the 5G SL software was verified by integrating the code into SDR platforms and undergoing OTA testing. The OTA radio tests show promising results that align well with the RFSIM results.

## II. 5G SIDELINK BACKGROUND

In this section, we introduce our added-on modules for 5G SL support, in OpenAirInterface5G (OAI). For this work, the focus was on the development and integration of two 5G SL Physical (PHY) layer channels, the PSBCH and PSSCH, for 5G SL mode 2. Ultimately, our goal is to have a complete SL architecture that can operate in both *mode 1*: UEs are within range of the base station's (gNB's) network coverage and SL resources are managed by the gNB; and *mode 2*: UEs operate without network coverage and can dynamically select their SL resources. Fig. 1 shows the SL modes and their components.

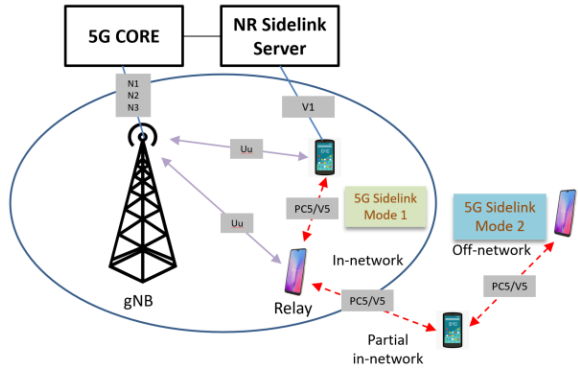

Fig. 1. 5G Sidelink Mode 1 and Mode 2 Configurations

### A. LTE Sidelink Versus 5G Sidelink

OAI successfully delivered LTE SL (ProSe) following 3GPP Rel. 14 36.211 to 36.214 specs in 2019. In LTE SL, the UE can operate in 4 modes: on-network coverage where UEs can connect to the eNB via the Uu interface, off-network coverage, or UE-to-UE via the PC5 interface, partial coverage UE-to-network relay procedure, and multiple node support or one-to-many direct communications.

LTE ProSe physical layer contains the following channels and signaling: Physical SL Control Channel (PSCCH), Shared Channel (PSSCH), Broadcast Channel (PSBCH), and SL primary and secondary synchronization signals (S-PSS and S-SSS). NR SL has the Physical SL Feedback Channel (PSFCH) for Hybrid ARQ(Automatic Repeat reQuest) or HARQ feedback transmission and reception, while LTE SL does not. Another key difference between LTE SL and 5G SL is the use cases that each technology can support. LTE SL uses SC-FDMA (Single-Carrier Frequency-Division Multiple Access) and supports 10 MHz and 20 MHz channels. The channel is divided into 180 kHz Resource Blocks (RBs), which correspond to 12 subcarriers of 15 kHz each. 5G SL on the other hand, uses OFDM with cyclic prefix (CP), which has a frame structure organized in radio frames, each with a duration of 10 ms. A radio frame is divided into 10 subframes, each with a duration of 1 ms.

The number of slots per subframe and the subcarrier spacing (SCS) for the OFDM waveform can be flexible for 5G SL. The scalable OFDM numerology configuration factor can be set as $\mu = 0,1,2,3$ such that the SCS can be equal to 15 kHz, 30 kHz, 60 kHz, or 120 kHz, respectively. 5G SL can also operate on frequency range 1 (FR1) and range 2 (FR2) with dedicated bands for SL [3]. Although the components of 5G and LTE SL physical layer are similar, their structures are different. In 5G SL, the flexible slot configuration allows for higher frequency ranges to be used, which will result in improved performance. When the wireless environment changes rapidly, the 5G time-domain format is adaptable to different capacities. The performance advantages include lower latency payloads and fast packet delivery.

### B. 5G Physical Layer Overview

When the wireless environment changes rapidly, the 5G time-domain format is adaptable to different capacities. In 5G, an RB is defined as a consecutive group of subcarriers, serving as the fundamental unit for allocating resources to users in the frequency domain [3]. Each RB is comprised of multiple subcarriers, known as resource elements (REs). These REs can be assigned to different elements for various purposes, such as PSBCH, PSSCH, and demodulation reference signals (DMRSs) for transmitting broadcast control data, user-plan data, and for channel estimation, respectively. This adaptable framework also increases performance for 5G communications.

## III. TECHNICAL APPROACH

### A. Additions to OAI

In this work, the development and test cycles were split based on the following two procedures.
- **Synchronization via the PSBCH:** The purpose of this development is to showcase the process in which the Synchronization Reference (SyncRef) UE can broadcast timing and frequency information. Any nearby UE, within range, can decode this information. The nearby UEs will update their own time and frequency states to enable communication between the SyncRef and other synchronized nearby UEs. This concept is denoted as *synchronization*. Various environments and configurations were tested to validate that the synchronization procedure is working and to understand the limitations of our prototype.
- **Data Transmission and Reception via the PSSCH:** The purpose of this development is to showcase the process in which the SyncRef UE can send data packets (text messages) to already-synchronized nearby UEs. The data reception provides opportunity for quality measurements to be captured which enables robust performance testing.

In this work, we developed the PSBCH and PSSCH in OAI's open-source repository (https://gitlab.eurecom.fr). The code is located on the "episys-sl-dev" branch as the team is actively working with OAI to merge the code to the "develop" branch. Although much of the 5G code is used as a baseline for the 5G SL development, many of the SL specific functions were built from scratch in conjunction with the 3GPP specification. Injecting 5G SL functionality into the existing OAI 5G protocol stack allows for faster releases to the community. The goal is to continuously integrate with OAI, to ensure the 5G SL development is designed with all OAI parallel efforts and bug fixes.

*B. Development and Test Cycle Description*

The current scope of development focuses primarily on the PHY layer. The MAC layer is lightly touched for pre-configuration of the sidelink master information block (SL-MIB). The development process utilizes existing 5G OAI code and the testing process includes robust OTA testing with various SDRs; specifically, the Universal Software Radio Peripheral (USRP) B210 and N310. The software development cycle consists of the following:
1. Research and study the corresponding 3GPP specification for the new channel implementation.
2. Modify and create the new 5G SL source code specific to the new channel.
3. Create a new OAI channel-specific simulator to conduct unit tests on the new channel implementation. These channel specific simulators can be used in CI/CD pipelines as the matured code is merged back to OAI.
4. Update the existing OAI full stack simulator (RFsimulator) to test the new channel with the full 5G protocol stack.
5. Integrate the new channel implementation with the current 5G SL development and test OTA via USRPs.
6. Repeat for next channel.

Fig. 2 illustrates both the simulator and OTA architecture. The simulation environment, denoted by the orange dotted line, validates the added 5G SL features prior to OTA testing. The final 5G SL prototype consists of three UEs conducting OTA 5G SL data transfer, utilizing the full 5G protocol stack higher layers, and the newly added 5G PHY channels.

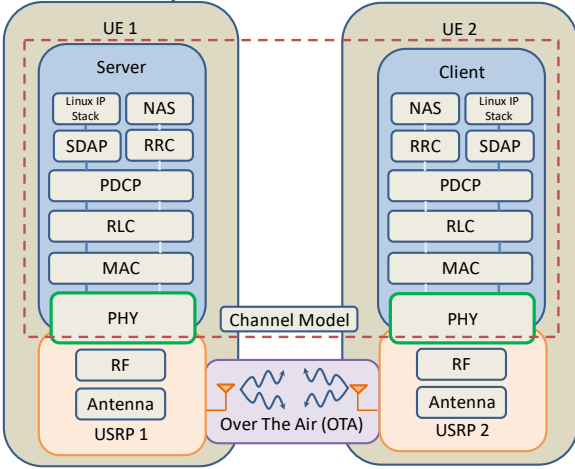

Fig. 2. 5G Sidelink Mode 2 Prototype and Simulation Architecture

## IV. SYNCHRONIZATION (PSBCH) IMPLEMENTATION

*A. Sidelink Synchronization Block (S-SSB)*

To establish a SL connection, synchronization information is broadcasted from a SyncRef UE and decoded by other nearby UEs. The combination of S-PSS and S-SSS is called the SL synchronization signal (SLSS) of the SL synchronization signal block (S-SSB). By detecting the SLSS, a nearby UE is able to extract critical pieces of information about the SyncRef UE to synchronize to it through estimation of the frame bounds and carrier frequency offsets.

The S-SSB occupies a time slot and uses the same numerology for general SL configurations. The S-SSB consists of PSBCH, S-PSS and S-SSS symbols. For a normal cyclic prefix (CP), an S-SSB slot consists of 14 symbols where symbol 0 is PSBCH with automatic gain control (AGC), symbols 1 and 2 are S-PSS, the next two symbols are S-SSS, symbols 5-12 are PSBCH, and the last symbol is reserved for guarding. In the frequency domain, the S-SSB uses 11 RBs, or 132 subcarriers. Fig. 3 shows the S-SSB slot with normal CP.

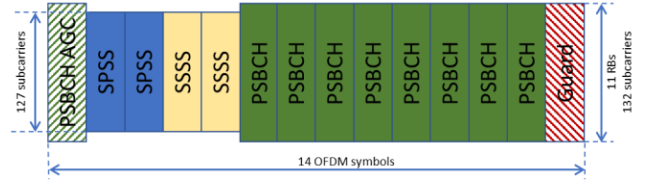

Fig. 3. 5G Sidelink S-SSB Slot Structure

The S-PSS and the S-SSS each consist of sequences of 127 bits. The S-PSS is generated from maximum length sequences (m-sequences) using the polynomial and initial values. There are only two candidate sequences are used for S-PSS. The S-SSS is generated from gold sequences. The SLSS ID or $N_{ID}^{SL}$ is identified as $N_{ID}^{SL} = N_{ID,1}^{SL} + 336 N_{ID,2}^{SL}$, where $N_{ID,1}^{SL} = \{0, 1, \dots, 335\}$ and $N_{ID,2}^{SL} = \{0, 1\}$ represent an identifier for S-SSS and S-PSS candidate sequences, respectively. S-PSS and S-SSS are modulated with BPSK and take 127 subcarriers each.

The PSBCH provides the system wide configuration and synchronization information required to establish the connection between multiple UEs. Its payload size consists of 56 bits, including the 1 bit in coverage indicator, 12 bits for indicating the time division duplexing (TDD) configuration, 10 bits for the direct frame number (DFN) and 7 bits for the slot index. The PSBCH payload also includes 2 reserve bits for future purposes as well 24 bits for a cyclic redundancy check (CRC).

*B. Synchronization Detection Process*

The receiver will first perform detection of S-PSS to extract $N_{ID,2}^{SL}$ using a cross-correlation algorithm in time domain, then taking the DFT and detecting S-SSS signals to extract $N_{ID,1}^{SL}$, in the frequency domain. The S-SSS symbols can be analyzed for RSRP measurements to report current channel conditions. After correlation, these signals are used to determine any frequency offset in the carrier frequency. The frequency offset is calculated by (1) below:

$$ssss_{fo} = \frac{4.3}{\pi} \times \theta, \quad (1)$$

where $\theta = \tan^{-1}\left(\frac{y}{x}\right)$, y and x are defined in (2) and (3).

$$y = \sum_{n=1}^{127} (d_{Nid1}(n) \times \text{Im}(sss_0(n)) + d_{Nid1}(n) \times \text{Im}(sss_1(n))), \quad (2)$$

and

$$x = \sum_{n=1}^{127} (d_{Nid1}(n) \times \text{Re}(sss_0(n)) + d_{Nid1}(n) \times \text{Re}(sss_1(n))) \quad (3)$$

In (2) and (3), $d_{Nid1}(n)$, is the reference sequence generated at the receiver.

Once the $N_{ID}^{SL}$ (composed of $N_{ID,1}^{SL}$ and $N_{ID,2}^{SL}$) is identified, the receiver starts to decode and extract data from the PSBCH payload. DMRS are transmitted in every PSBCH, on every fourth subcarrier; they are used by the receiver to properly decode the PSBCH. The payload will contain the current frame and slot information, which is used by the nearby UE to set these values accordingly. This will inherently synchronize the nearby UE timing. This procedure is depicted in Fig. 4.

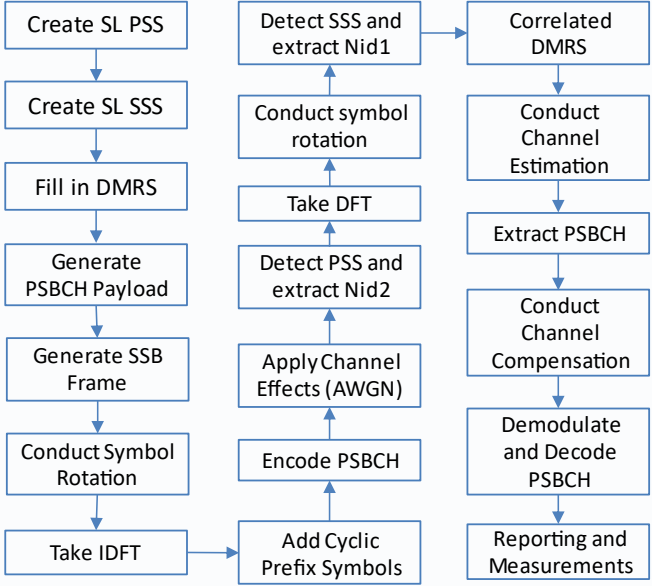

Fig. 4. 5G Sidelink S-SSB Synchronization Procedure Block Diagram

### C. TDD Slot Configuration

The flexible slot feature of 5G SL utilizes each symbol to schedule uplink, downlink, or flexible packets. The Radio Resource Control (RRC) layer of the protocol stack will apply the specific slot configuration and format to be used; the configuration is defined by the SyncRef UE and broadcasted in the first system information block (SIB1) to each nearby UE [6].

Given that the MAC layer is yet to be developed, the slot configuration was hardcoded in this work. The current modifiable parameters are shown in Fig. 5. The numbered items (2): Interval Parameter, (4): Offset Parameter, and (5): Number S-SSB per Period, are configurable. In a 30kHz SCS setup, the number of S-SSBs can be 1 or 2. According to the 3GPP standard and [6], the S-SSB should be periodically sent every 16 frames to achieve a periodicity of 160 ms (10 ms per system frame number (SFN), 0.5 ms per slot) in FR1.

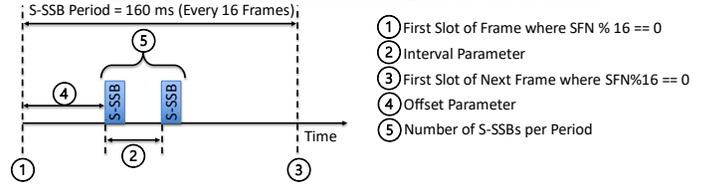

| Frequency Range | Subcarrier Spacing (SCS) | Number of S-SSBs per period |
|---|---|---|
| FR1 | 15 kHz | 1 |
| | 30 kHz | 1, 2 |
| | 60 kHz | 1, 2, 4 |
| FR2 | 60 kHz | 1, 2, 4, 8, 16, 32 |
| | 120 kHz | 1, 2, 4, 8, 16, 32, 64 |

① First Slot of Frame where SFN % 16 == 0
② Interval Parameter
③ First Slot of Next Frame where SFN%16 == 0
④ Offset Parameter
⑤ Number of S-SSBs per Period

Fig. 5. TDD Slot Configuration and S-SSB Periodicity

We arbitrarily chose to broadcast 2 S-SSBs per period. The interval parameter, (2), indicates the distance between the start of multiple S-SSB transmissions within the period. The offset parameter, (4), will determine the time offset (slot) within the eligible frame (must be a multiple of 16), that the first S-SSB in the period will be transmitted. In our case, the first S-SSB is sent in slot 2 because we set the offset parameter to 2. The interval parameter was set to 20; therefore, since the S-SSB will be transmitted twice in a period, the first transmission occurs in frame 0 slot 2, and the next occurs 20 slots later, in frame 1 slot 2.

## V. DATA TRANSFER (PSSCH) IMPLEMENTATION

### A. Physical Sidelink Shared Channel (PSSCH)

The development of the PSSCH is required for 5G SL data transmission and reception. The data passed across the PSSCH consists of 1st-stage SL control information (SCI1), and shared channel symbols (2nd-stage-SCI (SCI2), DMRS, and the payload). The SCI1 is used for the scheduling of PSSCH and SCI2. Fig. 6 shows a typical 5G SL slot without PSFCH, although the PSFCH may accompany with the PSSCH. From Fig. 6, the first symbol is used for AGC, which is simply a duplication of the next OFDM symbol. The next 3 OFDM symbols have partial PSCCH and PSSCH data. The mapping order of the modulation symbols for PSSCH DM-RS, SCI2, and data, to their respective resource elements (REs) is as follows: (a) PSSCH DM-RS symbols are mapped, (b) SCI2 symbols are mapped, and lastly, (c) data symbols are mapped.

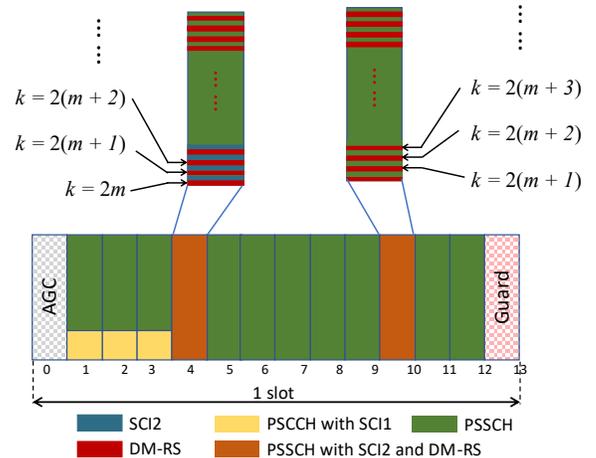

Fig. 6. NR Sidelink PSSCH Slot w/o the Physical Shared Feedback Channel

The SCI2 payload is polar encoded and contains the number of data bits transmitted as well as the multiplexing scheme; therefore, the SCI2 is used by the decoder to properly decode the PSSCH payload, which uses a Low-Density Parity Check (LDPC) encoding mechanism. The shared channel works in conjunction with the PSFCH to provide acknowledgements and channel information to the higher layers of the stack. Per section 8.4 of [7], the SCI2 is used for the decoding of PSSCH for the HARQ operation:

- HARQ-ACK information includes ACK or NACK,
- HARQ-ACK information includes only NACK,
- no feedback of HARQ-ACK information.

Among two SCI2 formats, SCI2-A is used for the decoding of PSSCH. The SCI2-A consists of 35 bits in total. There are 4 bits for the HARQ process ID, 1 bit for the new data indicator (NDI), 2 bits for the redundancy version, 8 bits for the source ID, 16 bits for the Destination ID, 1 bit for the HARQ feedback enabled/disabled indicator, 2 bits for the cast type indicator, and 1 bit for the CSI request. The upper layer RRC and MAC will fill out this information and form the SCI2 before transmission.

To determine the indices of the OFDM symbols that contain PSSCH DM-RS, Table 8.4.1.1.2-1 in [8] is used. Symbols 4 and 10 are assigned with PSSCH and DM-RS modulation symbols. Within each of these OFDM symbols, the modulation symbols for PSSCH DM-RS are allocated to the REs with $k = 2m$, where $m$ takes on values 0, 1…$n$, and $k$ does not surpass the total number of allocated REs. The modulation symbols for SCI2 are assigned to the available REs with $k = 2m + 1$ starting from the lowest index within the first OFDM symbol that contains PSSCH DM-RS. If there is a need for additional REs, the vacant REs in the subsequent OFDM symbol(s) will be used.

*B. PSSCH Implementaion*

After the synchronization process, a TX UE carries out a medium access algorithm in the MAC layer and starts transmission in PHY via the PSCCH and PSSCH. In some cases, the PSFCH is associated with the combined PSSCH/PSCCH slot as well. For simplification, we constructed the data frame with PSSCH and PSCCH only.

The control channel payload contains the SCI1. Since the primary focus of the development consists of the PSBCH and PSSCH, the development of the SCI1 is reserved for future work and was hardcoded. By fixing the SCI1 payload, an assumption was made that the receiver UE can successfully decode the required parameters (for properly decoding the SCI2 and payload), and this SCI1 data are passed directly to the receiver.

The second part of the PSSCH is the payload data. PSSCH is the main channel for transmitting user data and control information between transmitters and receivers. The PSSCH operates in the time and frequency domain and supports various modulation schemes determined by the modulation and coding scheme (MCS) index. It is important to note that the MCS is transmitted on the PSCCH, but in this work, it was hardcoded (and varied for performance tests) in the PHY layer. The components in the shared channel include cyclic redundant code (CRC), channel encoding/decoding, rate matching, data and control multiplexing, scrambling/descrambling, modulation/demodulation, and RB mapping. The SCI2 is encoded with polar coding and modulated as QPSK, while the payload is encoded with LDPC coding and modulated following the MCS indexing to maximize the capacity. The PSSCH transmission and reception procedure is summarized in Fig. 7.

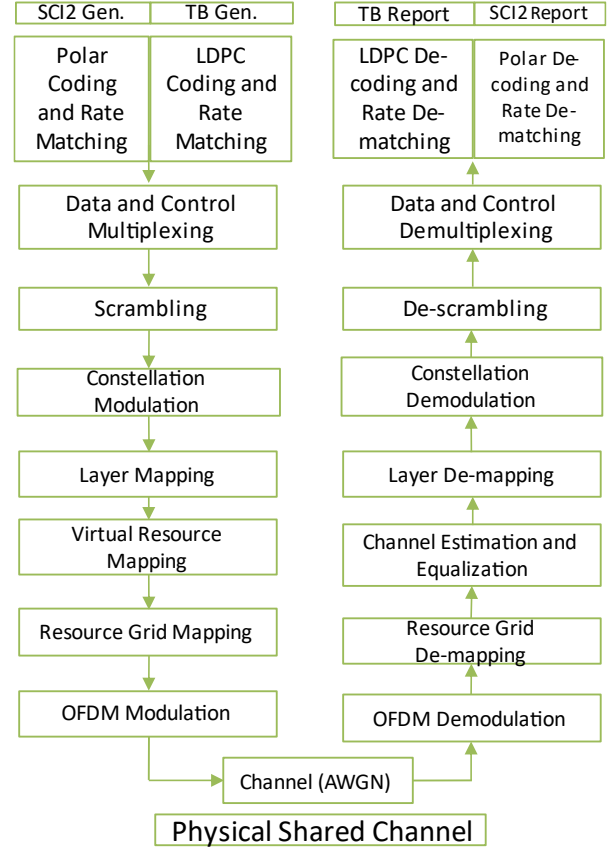

Fig. 7. PSSCH Transmission and Reception Procedure

## VI. TEST SETUP AND RESULTS

*A. Testbed and Equipment Description*

The full system test includes launching one OAI UE on a USRP, which will broadcast S-SSB frames periodically. The S-SSB frames are then received and decoded by another nearby USRP. In this work, we tested two and three-UE scenarios. The 3-UE setup consists of three UEs (three USRP B210s), where one USRP is the SyncRef UE, and the other two UEs are deployed as nearby UEs. Fig. 8 depicts our setup. Internally, we validated that multiple UEs are able to properly get sync and decode the PSSCH payloads. The performance tests in the following sections were conducted only for a 2 UE scenario.

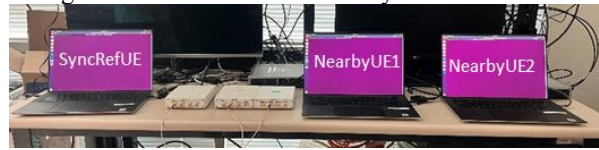

Fig. 8. 5G Sidelink Testbed Setup #1: Three B210 USRPs; One as the SyncRef Transmitting UE, and Two as Nearby Receiving UEs

The USRP B210 SDR is FPGA-based and equipped with a Spartan6 FPGA chip and is combined with AD9361 RFIC. Its frequency operation covers from 70MHz to 6GHz. The B210 is full duplexing, with MIMO (2Tx and 2 Rx) support, and can operate up to 56 MHz of real-time bandwidth (61.44MS/s quadrature). These B210s are connected with three Linux-based laptops via USB3.0. Ubuntu 20.04 with the 5.15.0-73-generic kernel was installed on these PCs. Each laptop is equipped with an Intel Core i7 10875H CPU@ 2.30GHz and 32GB of memory. The latest open-source USRP Hardware Driver (UHD) version 4.4.0 and FPGA binary firmware are used. The specific RF parameter configurations and MAC layer pre-configurations are shown in TABLE I.

TABLE I. RF PARAMETER CONFIGURATIONS AND SETTINGS

| Parameters | Value | Range |
|---|---|---|
| Tx Gain | 0 dB | {0, 100} dB |
| Rx gain | 20 dB | (0, 100} dB |
| # of tx antennas | 1 | {1} |
| # of rx antennas | 1 | {1} |
| Carrier frequency | 2600 MHz (n38) | {2600, 5900} MHz |
| Carrier bandwidth | 50 PRBs (18 MHz) | {50, 100} PRBs |
| Sampling rate | 30.720 MS/s | {7.68, 61.44} MS/s |
| Numerology index | 1 | {0, 1, 2, 3} |
| FFT size | 1024 Samples | {1024, 1536, 2048} |
| Cyclic prefix | Normal | {Normal, Extended} |
| PSSCH modulation | QPSK | {QPSK, 16QAM, 64QAM, 256QAM} |
| MCS | 0-9 | {0-31} |
| sl-NumSubChannel | 1 | Number of subchannels in the corresponding resource pool |
| sl-SubchannelSize | 50 | Minimum granularity in frequency domain for the sensing for PSSCH resource selection |
| sl-StartRB-Subchannel | 0 | Lowest Resource Block index of the subchannel with the lowest index in the resource pool |
| sl-TimeResource | FFFFF | Bitmap of the resource pool |
| sl-LengthSymbols | 14 | Number of symbols used for sidelink in a slot without S-SSB |
| sl-TDD-Config | Default | 11 bits of Time Division Duplexing |
| directFrameNumber | 2 | Frame number in which S-SSB transmitted |
| slotIndex | 2 | Slot index in which S-SSB transmitted |

*B. RF Simulator (RFSIM) Results*

Prior to conducting OTA testing via SDRs, we collected system performance data through RFSIM by generating BLER curves for various SNR levels and MCS indices. For simplicity, the simulation test was conducted for a 2 UE scenario. This simulator allows the OAI full stack to be tested without an RF board and replaces the actual RF board driver. Therefore, it functions like an RF board but not in real time; the simulator is CPU bound. In other words, we simulate the synchronization and decoding procedure between two devices without the presence of actual RF modulation and demodulation. Fig. 9 shows the basic RFSIM architecture.

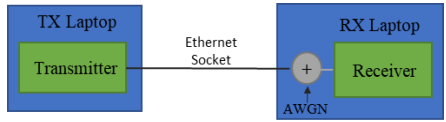

Fig. 9. RF-Simulator Setup

Fig. 10 shows the BLER performance of the PSSCH decoder for ten MCS values: $I_{MCS} = 0, 1, 2, 3, 4, 5, 6, 7, 8$ and $9$. We conducted PSSCH data traffic in RFSIM as a function of SNR in an AWGN channel. After synchronization, the SyncRef UE will transmit data packets, where the transport block size (TBS), code block size (CBS) and segmentation of the packet is dependent on the MCS used. The determination of the BLER value was based on the number of properly LDPC-decoded packets over the total number of received packets (4).

$$BLER = 1 - \frac{\# \, of \, Decoded \, Packets}{\# \, of \, Received \, Packets}. \quad (4)$$

The log files reporting the number of decoded packets and received data packets were analyzed and the data was ported to MATLAB for plotting. Note, this figure shows the performance of the decoder in the presence of time synchronization but without conducting carrier frequency offsetting (CFO).

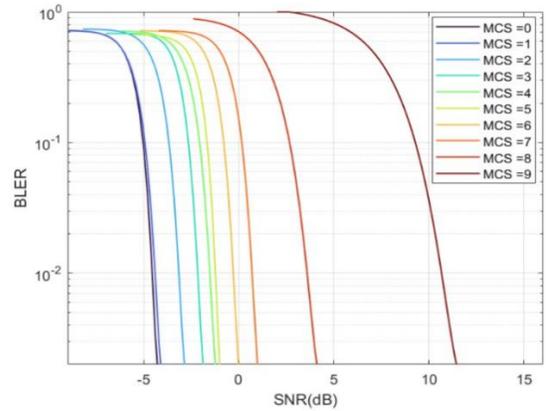

Fig. 10. RF-Simulator BLER versus SNR for various MCS settings. 10 trials for each MCS and SNR value   Over-the-air (OTA) USRP Results

*1)   OTA Synchronization (PSBCH) Results*

The first benchmark of our OTA test setup is the synchronization procedure. The SyncRef UE sends the S-SSB data periodically and we capture the number of successfully decoded S-SSBs at the nearby UEs and record the SSB reference signal received power (RSRP) values. The RSRP is calculated as the average power contribution of the resource elements that carry the reference signals. It is measured based on PSBCH DMRS, PSCCH DMRS, and PSSCH DMRS independently. For our experiments, we have calculated two RSRP values, one for the S-SSB and for the PSSCH, namely SSB-RSRP and PSSCH-RSRP, respectively. For the synchronization validation test, the SSB-RSRP was analyzed.

To create a controlled channel environment, we artificially vary the channel conditions by adding an attenuator (Mini-Circuits Model RC4DAT-8G-120H, range of 0-120dB with 0.05 dB step size) to the nearby UEs. As shown in TABLE II. when we increase the attenuation from 0 to 25dB, the SSB-RSRP decreases; accordingly, this decrease indicates the received signal power is lower. When the signal power is lower (e.g., attenuation is higher), the number of properly received packets decreases. This expected trend, a decrease in successful S-SSB receptions as the channel conditions are becoming more adverse (e.g., attenuation is higher), was used to validate the synchronization procedure. Once we completed the synchronization validation, we began to conduct performance and verification tests via OTA data traffic across the PSSCH.

TABLE II. OTA SYNCHRONIZATION TEST WITH USRP B210S

| Attenuation (dB) | SSSB-RSRP (dBm) | # of Received Packets |
|---|---|---|
| 0 | -60.8 | 1105 |
| 5 | -60.57 | 1111 |
| 10 | -68.83 | 1068 |
| 15 | -62 | 673 |
| 20 | -67 | 668 |
| 25 | -72 | 679 |

*2) OTA Data Payload (PSSCH) Results*

In the second portion of the OTA testing, after successfully transmitting the S-SSB data, the SyncRef UE starts to transmit PSSCH data. PSSCH data is user data and is described in detail in previous sections. In order to validate the OTA performance, we conducted a proof-of-concept distance test, between 2 N310 USRPs. The setup of this test involved setting the TX UE a variable distance from the RX UE. The SyncRef (TX) UE would periodically transmit user data (periodicity was determined in section IV.C). The data was then received, and the successful decoding rate was recorded. The BLER value was calculated based on equation (4).

In the case of OTA testing, calculating the *# of Received Packets* is not as simple as interpreting the RX logs because the USRP RF front end is always listening, whether data is currently being transmitted or not. For example, if the SyncRef has not started to transmit data, the RX UE would retrieve noise and report packet failures. To avoid false negatives, in this test, we always started the transmitter first, to ensure that data was available to reception. Furthermore, we did not consider packet loss prior to synchronization, to ensure the connection between the SyncRef UE and nearby UE was stable and possible.

We repeated the test for five runs at nine various distances (1, 2, 3, 4, 5, 6, 7, 8 and 9ft between the SyncRef and Nearby UEs) and averaged the BLER vs (measured) SNR at each location. For this test the MCS value was set to 9. Moreover, we varied the max number of LDPC iterations. The number of LDPC iterations will improve the RX decoding ability but will trade throughput and latency performance [9]. One of the drawbacks of the BLER vs. Distance test is the environment limitations found in the lab. These include variable channel conditions caused by changes in weather, time of day, Wi-Fi interference, other nearby device interference, and more. This test was conducted as a proof-of-concept; however, the team is aware that the results can easily vary depending on external factors. Due to these imperfections, the purpose of this test was to show that there will be a specific threshold in which the USRPs will not be able to communicate – in our lab scenario this was around 8-9ft. To definitively test the bounds of the system, one would need a controlled environment, such as an anechoic chamber. The distance test setup is shown in Fig. 11.

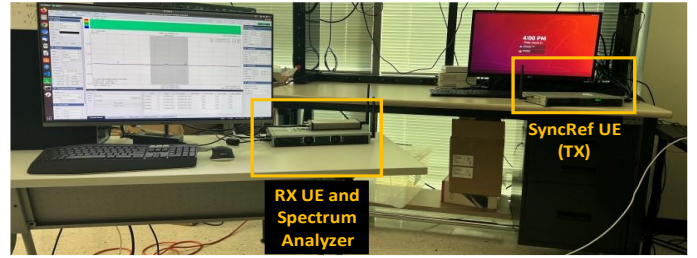

Fig. 11. BLER vs. Distance Test Setup (with nearby UE, Spectrum Analyzer, and SyncRef TX UE)

For the BLER vs. Distance test, a spectrum analyzer was placed at the same distance as the RX UE and would measure the RX power and noise. These values were then used to calculate the approximate SNR of the received packets, see equations (5, 6).

$$SNR = 10 log_{10}\left(\frac{RX\ Power_{Watts}}{RX\ Noise_{Watts}}\right) \quad (5)$$

$$RX\ Power_{watts} = 1_{watt}\frac{10^{RX\ Power_{dBm}/10}}{1000} \quad (6)$$

$$= 10^{(RX\ Power_{dBm}-30)/10}.$$

Once the SNR value was measured and calculated, we plotted the BLER vs. SNR, which is shown below in Fig. 12, where each SNR value corresponds to the varying distance set {1, 2, 3, 4, 5, 6, 7, 8, 9} ft. The BLER vs. SNR results show a promising behavior; as the SNR is increased, the BLER value is decreasing. As aforementioned, we ran this test 5 times; with hundreds of thousands of runs, the results would likely show much smoother BLER curves; yet again, this test was conducted as a proof-of-concept test. Furthermore, the environment of the lab does not allow for a controlled SNR measurement.

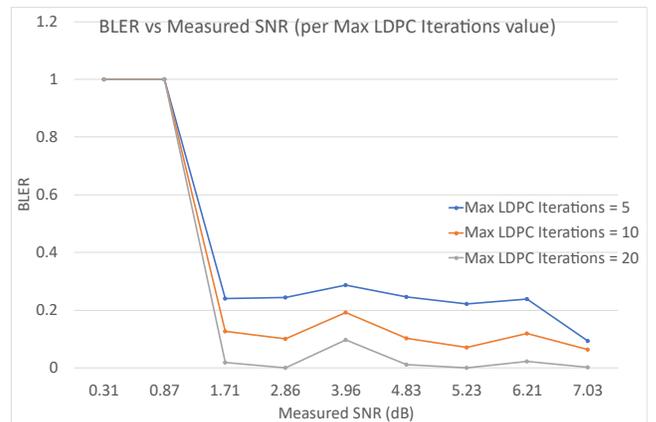

Fig. 12. BLER vs. Measured (Approximated) SNR Results (2 UE scenario)

One of the drawbacks of the BLER vs. Distance test is the environment imperfections found in the lab. However, the RFSIM test results give us confidence in the system as the controlled experimental results are consistent with the SNR and block error rate relationship.

## VII. CONCLUSION

In conclusion, this work was designed to accomplish the following.
- Provide an in-depth account of the procedures and developed source code for a Rel. 16 3GPP compliant 5G SL prototype. This includes details of both the implemented PSBCH and PSSCH in the OAI repository.
- Analyze the performance of the newly developed 5G SL (mode 2) prototype, including BLER curve reporting in simulation and OTA environments. It also includes a controlled experiment conducted by varying attenuation levels to support proof of concept of the PSBCH.
- Discuss the test set up and protype environment. Our lab has the ability to test with 3 B210 USRPs and 2 N310 USRPs; all performance metrics were tested with 2 UE scenarios. Documentation of the configurable parameters of the system and the UE settings are provided herein.
- Report any shortcomings and assumptions made in the development process that are to be addressed in the near future, including higher layer development.

Ultimately, the RFSIM results were compared against the OTA USRP testing. The results are shown below in Fig. 13. These results indicate that the PSBCH and PSSCH implementation was robust and able to perform as expected in both a simulated and OTA environment. In this experiment, we overlayed the RFSIM and OTA data. The BLER results of MCS 9 from the RFSIM were used with the OTA max LDPC iterations = 10 BLER results, while using a cubic smoothing spline on the OTA data. The results shown indicate that the overall performance is consistent; once the SNR is above 4 dB, the BLER drop significantly. However, in RFSIM this occurs around 4 dB and based on Fig. 13, this occurs between 5-6 dB OTA; it is also likely due to environmental interference and a small number of trials conducted OTA. The SNR values between 0 dB and 2 dB in the OTA environment have a large drop off. This drop off is attributed to the 8-9 ft distance in which the UEs were unable to communicate, then as the UEs were moved closer, the BLER values quickly rapidly. This is also shown in Fig. 12 w.r.t. the measured SNR at these distances. A smaller step size in SNR (or distance) and more trials would likely smooth this transition.

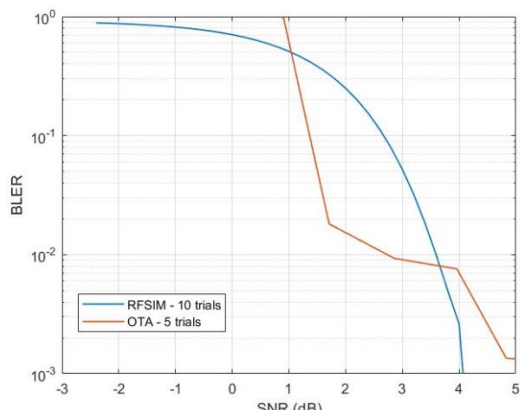

Fig. 13. RFSIM BLER vs. OTA Data Transmission BLER

## VIII. FUTURE WORK

### A. Physical Sidelink Feedback/Control Channel (PSFCH/PSCCH) and Higher Layer Implementation

The PSFCH can provide HARQ feedback information to a PSSCH reception from RX UE(s) to TX UE. PSCCH provides information about the MCS, power control, and other parameters that are necessary for establishing reception of the PSSCH. As part of our plan to enable end-to-end communication on 5G SL stack, the Radio Resource Control (RRC) protocol and MAC layer protocol will be developed next.

### B. Sensing Procedure/Resource Pool Allocation

When a UE needs to send traffic, the UE will set a sensing window and selection window for the resource selection. During the sensing window, the UE measures the RSRP of DMRS on the PSSCH (or PSCCH) of all the considered subchannels. During the selection window, the UE can select subchannels among the candidate resources in a random order or a predetermined FIFO order. The subchannel selection method, random or FIFO, is determined at the time of implementation, and is an area where machine learning techniques can be utilized to provide optimized resource allocations and selections.